\documentclass[useAMS,usenatbib]{mn2e}
\usepackage{natbib}
\usepackage{graphicx}

\title[First stars polluting $\omega$~Cen]{First stars as a possible origin for the helium-rich population in $\omega$~Cen}
\author[E. Choi \& S. K. Yi]{Ena Choi and Sukyoung K. Yi \\
Department of Astronomy, Institute of Earth Atmosphere Astronomy, Yonsei University, Seoul 120-749, Korea}

\begin{document}

\pagerange{\pageref{firstpage}--\pageref{lastpage}} \pubyear{2006}

\maketitle

\label{firstpage}

\begin{abstract}
The most massive Galactic globular cluster $\omega$~Cen appears to have
two, or perhaps more, distinct main sequences. Its bluest main sequence is
at the centre of debate because it has been suggested to have an extremely
high helium abundance of $Y \sim 0.4$. The same helium abundance is claimed
to explain the presence of extreme horizontal branch stars of $\omega$~Cen
as well. This demands a relative helium to metal enrichment of
$\Delta$$Y$/$\Delta$$Z$ $\sim 70$; that is, more than one order of magnitude
larger than the generally accepted value.
Candidate solutions, namely, AGB stars, massive stars, and supernovae,
have been suggested; but in this study, we show that none of them is
a viable channel, in terms of reproducing the high value of
$\Delta$$Y$/$\Delta$$Z$ for the constrained age difference between
the red and blue populations. 
Essentially no populations with an ordinary initial mass function, 
including those candidates, can produce such a high $\Delta$$Y$/$\Delta$$Z$
{\em because they all produce metals as well as helium}.
As an alternative, we investigate the possibility of the stochastic
``first star'' contamination to the gas from
which the younger generation of $\omega$~Cen formed.
This requires the assumption that Population III star
formation episode overlaps with that of Population II.
While the required condition appears extreme, very massive objects in
the first star generation provide a solution that is at least as 
plausible as any other suggestions made before.
\end{abstract}


\section{Introduction}
The most massive Galactic globular cluster, $\omega$~Cen,
shows a double (or more) main sequence in the colour-magnitude diagrams (CMD),
with a minority population of the bluer main sequence ($bMS$) and
a majority population of redder MS ($rMS$) stars 
(Anderson 1997; Bedin et al. 2004).
The interpretation of the $bMS$ by a huge excess in the helium abundance
($Y \sim 0.4$) has been suggested via CMD fittings (Norris~ 2004) and
supported by the spectroscopic study on the $bMS$ stars (Piotto et al.~ 2005).
The extreme helium abundance is claimed to
explain the extremely-hot horizontal-branch (HB) stars of
$\omega$~Cen as well (Lee et al. 2005).

The helium-enrichment parameter 
$\Delta$$Y$/$\Delta$$Z$ required to explain the sub-population of
$\omega$~Cen are $\sim 70$, more than an order of
magnitude larger than the currently-accepted value, 1--5 
(Fernandes et al.~1996; Pagel \& Portinari 1998; Jimenez~et~al.~2003).  
To explain this, several possible polluters,
i.e., AGB stars, stellar winds associated with massive stars during their
early evolutionary phases and type II supernovae (SNe II) have been suggested
(e.g. Norris 2004; D'Antona et al 2005).
However, Bekki \& Norris (2006) show that these candidates cannot reproduce
such a high {\em helium abundance} for reasonable initial mass functions
(IMFs) within the scheme of a closed-box self-enrichment.

The acute point is not the high helium abundance itself
but the high value of $\Delta$$Y$/$\Delta$$Z$, because $populations$
containing those candidates produce a large amount of metals as 
well as helium according to canonical stellar evolution models.
Maeder \& Meynet (2006) have recently suggested that ``moderately'' 
fast-rotating massive ($\sim 60~M_\odot$) stars expel helium-dominant 
gas into space and thus could provide a solution to this problem. 
We show that this scenario provides a viable solution only if
additional conditions on the mixing and escaping of metals are invoked.

As an alternative solution, we investigate the possibility that
the late generation of ``the first stars'' (Population III)
might affect the chemical mixture of the minority of Population II stars.
We are particularly inspired by the work of Bromm \& Loeb (2006)
that suggested an overlap between the Population III and II star
formation episodes. The chemical yields of such supposedly-heavy,
zero-metal stars have been computed by Marigo, Chiosi, \& Kudritzki (2003).
We present in this Letter the result of our investigation.

\section{Chemical Evolution Models}

In order to investigate the pattern of the helium enhancement in a population,
we employ a simple chemical evolution code and a realistic initial mass
function (IMF).
We describe a simple chemical enrichment model essentially following the
formalism of Tinsley (1980) and reduce a set of a few parameters
following the formalism of Ferreras \& Silk (2000, 2001).
A two-component system is considered, consisting of cold gas and
stellar mass. The net metallicity $Z$ and helium content $Y$ of two systems
are traced. We assume instantaneous mixing of the gas ejecta from
stars and instantaneous cooling of the hot gas component.
The mass in stars,
$M_s(t)$ and in cold gas, $M_g(t)$, are normalised to the initial gas mass,
\begin{equation}
\mu_s(t)\equiv {M_s(t)\over M_g(0)},~~~\mu_g(t)\equiv {M_g(t)\over M_g(0)}
\end{equation}
where the initial states of the galaxy is assumed to be completely gaseous
without stars: $\mu_s(0) = 0$.

A Schmidt-type star formation law (Schmidt 1963) is assumed:
\begin{equation}
\psi(t)=C_{\rm eff} M_g^n(t)
\end{equation}
where the parameter $C_{\rm eff}$ implies the star formation efficiency.
We assume $n=1$, that is, a linear law $\psi(t)=C_{\rm eff} M_g(t)$.

Exponential infall of primordial gas is assumed:
\begin{equation}
f(t) = \Theta(t-\tau_{lag}) A_{inf} e^{-(t-\tau_{lag})/\tau_{inf}}
\end{equation}
where, $\Theta(t)$ is a step function.
The parameters $A_{inf}$, $\tau_{inf}$, and $\tau_{lag}$ are
the infall rate, timescale, and delay, respectively.
In order to explain the G-dwarf problem (Tinsley 1980),
gas infall has been considered as a solution (Larson 1972).

Some cold gas is heated to high temperature by supernovae
and/or AGN and can be driven out: outflows.
It is also an important factor to the final chemical properties of 
galaxies (Larson 1974; Arimoto \& Yoshii 1987).
We use a free parameter $B_{out}$ that represents the
fraction of gas ejected following the formalism of
Ferreras \& Silk (2000). This parameter should be a function of the mass
of the galaxy, whose potential well determines whether the winds are strong
enough to escape its gravitational potential.

There are five input parameters : $C_{\rm eff}$, $B_{out}$, $A_{inf}$,
$\tau_{inf}$, $\tau_{lag}$ with the following initial conditions:
the initial metallicity  $Z_0=10^{-4}$ and the initial helium
of stars and gas $Y_0 = 0.235$, the IMF slopes and
cutoffs.

\subsection{Mass Evolution}
The evolution of gas mass is given by
\begin{equation}
\begin{array}{ll}
{d\mu_g\over dt}& = (1-B_{out})E(t)-C_{\rm eff}\mu_g(t)\\
 & \\
 & +\Theta (t-\tau_{lag})A_{inf}e^{-(t-\tau_{lag})/\tau_{inf}}
\end{array}
\end{equation}
where the stellar mass ejecta $E(t)$ is defined as
\begin{equation}
E(t) = \int_{m_t}^\infty dm\phi (m)(m-w_m)C_{\rm eff}\mu_g(t-\tau_m).
\end{equation}
where $\phi(m)$ is the Scalo IMF 
(Scalo 1986) with cutoffs at 0.1 and 100$M_{\odot}$.
We adopt the broken-power law of Ferreras \& Silk (2000, Eq. 6) for the
stellar lifetime. Remnant mass for a star with main sequence mass
$m$, $w_{m}$, is adopted from Ferreras \& Silk (2000).

\subsection{Chemical Evolution}
The chemical evolution of gas is given by
\begin{equation}
\begin{array}{ll}
{d(Z_g\mu_g)\over dt}& =-C_{\rm eff}Z_g(t)\mu_g(t)+(1-B_{out})E_Z(t)\\
 & \\
 & +\Theta (t-\tau_{lag})Z_fA_{inf}e^{-(t-\tau_{lag})/\tau_{inf}}
\end{array}
\end{equation}
\begin{equation}
\begin{array}{ll}
E_Z(t)& =\int_{m_t}^\infty dm\phi (m)C_{\rm eff}\big[ (m-w_m)
        (Z_g\mu_g)(t-\tau_m) \\
 & \\
 & +mp_m\mu_g(t-\tau_m)\big],\\
\end{array}
\end{equation}
where $p_m$ denotes the mass fraction of a star of mass $m$ that is $newly$
converted to metals or helium and ejected. We approximate it
by a polynomial fit to the $p_m$ prediction of Maeder (1992).
In Figure 1, we show Maeder's chemical yields for metal-poor ($Z=0.001$) stars
and our fitting functions.
For our reference model, we use these chemical yields. 

We also show the metal yields of extremely metal-poor ($Z=0.00001$) 
rotating stars (Maeder \& Meynet 2006) in dotted line.
Maeder \& Meynet computed the yields for $60~M_\odot$ stars, and
in order to domonstrate their effects to the $\Delta$$Y$/$\Delta$$Z$ 
of an integrated population, we set up an extreme assumption that the 
newly-proposed metal yields may be applicable to all heavy stars of $M\geq 50$.

The crucial point shown here is that canonical stellar models 
{\em produce and spread into space metals as well as helium.}
No combination of these stars can produce $\Delta$$Y$/$\Delta$$Z$ that 
exceeds the values of the constituent stars.
The rotating metal-poor massive star models by Maeder \& Meynet (2006)
are exception and indeed can produce high values of $\Delta$$Y$/$\Delta$$Z$.
So we explore below whether a $population$ including such stars
can indeed present a solution to our problem.

\begin{figure}
\begin{center}
\includegraphics[width=8cm]{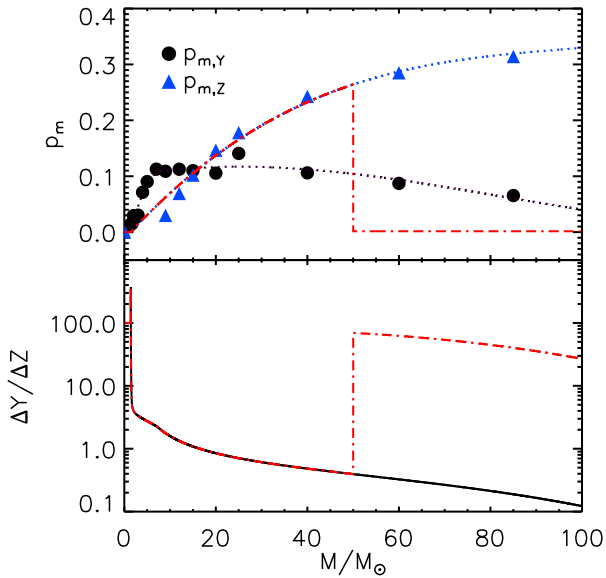}
\caption{
$top$: The chemical yields of metal-poor ($Z=0.001$) stars from 
Maeder (1992) and our fits to them. The red dash-dot line is for our ad-hoc 
case employing the $Z=0.00001$ rotating massive models by Maeder \&
Meynet (2006) for $M\geq50 M_\odot$. 
$bottom$: The resulting helium-to-metal ratios (bottom). The red line
is for the ``ad-hoc case''.}\label{fig:fig1}
\end{center}
\end{figure}

SNe Ia are generally considered
important in chemical evolution studies because they eject a considerable
amount of iron-peak elements: $\sim 0.7~M_{\odot}$ (Tsujimoto et al. 1995).
For our exercise, we have tried the most up-to-date SN Ia rate of
Scannapieco \& Bildsten (2005), consisting of two components:
star formation rate and the total stellar mass of the system.
The contribution by SNe Ia is negative to our study because
they are copious producers of metals. We remove its effects for simplicity.

We have found that gas infall or outflows of the processed material 
have negligible impacts on $\Delta$$Y$/$\Delta$$Z$ unless specific chemical
elements alone (e.g., helium or metals) are affected. 
Hence, only the closed-box model is considered.

The star formation efficiency parameter $C_{\rm eff}$ has a small effect.
When a range $C_{\rm eff}$ = 0.1 -- 5.0 $\rm Gyr^{-1}$ is considered,
a population can reach $\Delta$$Y$/$\Delta$$Z \approx 1$ -- 2 after
a 1~Gyr evolution. We show a typical case of $C_{\rm eff}=1$ in Figure 2
as a solid line.
The value of our reference model
is consistent with the most recent observational
mean value of the helium enrichment parameter of the galactic gas.

Having failed to recover the high $\Delta$$Y$/$\Delta$$Z$ in the
standard chemical enrichment model, we consider the stochastic effect
in the helium yields $p_{m,Y}$.
We assume that the stars in a specific mass range can have larger
values of $p_{m,Y}$ than the current stellar physics suggest.

Fig. 2 presents part of the result of the stochastic-effect test,
where $C_{\rm eff}=1$ is assumed;
(1) when $p_{m, Y}$ is doubled for all the stars,
(2) when $p_{m, Y}$ is doubled for 6 -- 10~ $M_{\odot}$ stars, maximising
the AGB effect, and
(3) when $p_{m, Y}$ is doubled for 40 -- 50~ $M_{\odot}$ stars,
allowing the SN II effect to dominate.
Lastly, the triple-dot dashed line is for the model where
all heavy stars ($M\geq 50$) are fast-rotating showing yields of Maeder
\& Meynet (2006). 
Because instant recycling of chemical elements is unrealistic, we use
the chemical mixing time scale of 50~Myr.

The $\Delta$$Y$/$\Delta$$Z$ after a 13 Gyr evolution of the AGB-dominant
system is merely $\sim~2$ times greater than that of the reference system.
The effect is even smaller for the SNe II dominant system.
Even if we double the values of $p_{m,Y}$ for all stars, the system could
attain only $\Delta$$Y$/$\Delta$$Z \approx4$.
But of course, for our purpose, we need a much higher helium enrichment within
a much shorter timescale ($\approx 1$~Gyr).

\begin{figure}
\begin{center}
\includegraphics[width=8cm]{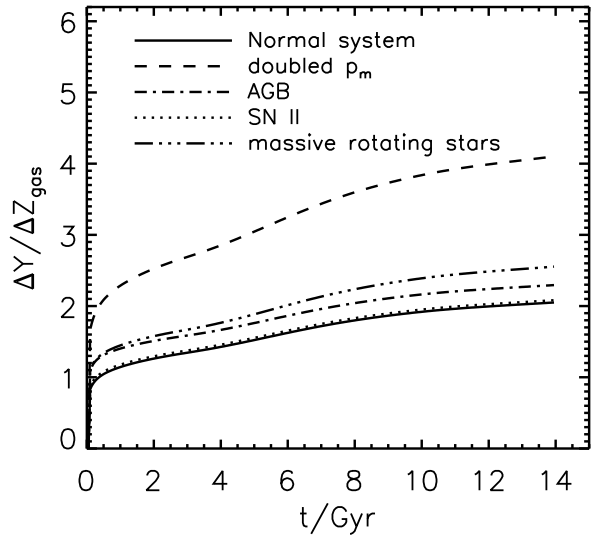}
\caption{The helium enrichment parameters resulting from the stochastic
effect. Our reference model is shown in solid line.
Other lines are for the cases that $p_{m,Y}$ is doubled for
(1) all stars, (2) 6 -- 10~$M_{\odot}$ stars (AGB progenitors) stars,
and (3) 40-50~ $M_{\odot}$ (SN II progenitors).
Finally, triple-dot dashed line shows the model that assumes that
all heavy stars ($M\geq 50$) are fast-rotating showing yields of Maeder
\& Meynet (2006). All models are for the case of $C_{\rm eff}=1$ and the
chemical mixing time scale of 50~Myr.
\label{fig2}}
\end{center}
\end{figure}

Doubling the helium yields is ad hoc and hard to
justify in terms of nucleosynthesis. Even in this unphysical test,
however, the stochastic effect does not help us achieve 
the high $\Delta$$Y$/$\Delta$$Z$.
This is because (1) ordinary stars produce and spread into space not
only helium but also metals and (2) most notably in the case of 
the ``Maeder \& Meynet case'', there are many more low-mass stars in
the ordinary IMF than massive stars. 

Other possible contributing factors, e.g., IMF slope and the remnant mass 
$w_m$, also have some influence. 
One may wonder about the effect of radically-different IMFs
(e.g., with a reversed slope). This indeed would produce more helium
but even more metals in ordinary stellar populations, hence in fact 
lowering the resulting value of $\Delta$$Y$/$\Delta$$Z$.
On the other hand, a reversed IMF slope would dramatically increase 
$\Delta$$Y$/$\Delta$$Z$ in the ``Maeder \& Meynet case''. However,
this should not be common because otherwise all populations
with massive stars should have enhanced helium. 

When assuming that the helium enrichment of $bMS$ of $\omega$~Cen is due to
the ejecta from the $rMS$ formed earlier, the candidate solution is even
more severely constrained. For example, low-mass stars which produce mainly
helium can generate high values of $\Delta$$Y$/$\Delta$$Z$ upto
$\sim 8$ but they take much too long to spread the processed materials
into space, when compared to the age difference between the two sub-systems
in $\omega$~Cen, i.e., 1--2~Gyr (Lee et al. 2005).
Any star whose lifetime is longer than this would have no impact at all.

The ``Maeder \& Meynet case'' is noteworthy as it is the only scenario
where high values of $\Delta$$Y$/$\Delta$$Z$ can be achieved during
a short period of time. Assuming all stars form simultaneously in a population,
$\Delta$$Y$/$\Delta$$Z$ can be as high as we need during the first 
few Myr until intermediate-mass stars begin to spread a large amount of
gas into space. In this sense, this scenario provides a solution
if the $bMS$ population of $\omega$~Cen is only slightly younger than
its $rMS$ population. However, the time window ($\sim 10^{7-8}$~yr) 
for this to happen is very small compared to the constraint from 
CMD studies (Lee et al. 2005). 
The possibility of difference in the effect of stellar wind for helium 
and metals demands further studies, and the role of the local black 
hole accretion on metals may also be noteworthy (Maeder \& Meynet 2006).

\section{Very Massive Objects}

\begin{figure}
\begin{center}
\includegraphics[width=8cm]{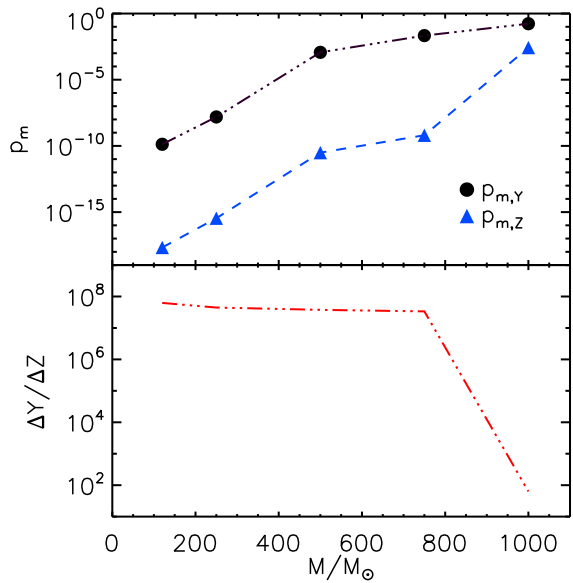}
\caption{ (Top) Filled circles represent the $p_{m,Y}$, $p_{m,Z}$, newly synthesized and ejected helium or metal
mass fraction, calculated from yields of Marigo et al. (2003) for 120, 250, 500, 750 and 1000 $M_{\odot}$ VMOs.
Dotted line implies linear interpolation of given 5 points in logarithmic scale. (Bottom) The $\Delta$Y/$\Delta$Z
of each mass of VMOs}
\end{center}
\end{figure}

We finally explore the plausibility of the ``first'' Pop III star pollution.
The first stars in the universe are thought to have formed out of the
metal-free gas at redshifts z$\ga$10 (Bromm~\&~Larson~2004;
Ciardi~\&~Ferrara~2005).
These Pop III stars are often predicted to be very
massive, $M\ga 100M_{\odot}$ (Bromm, Coppi \& Larson, 2002;
Abel, Bryan \& Norman 2002), although a different suggestion has also been
made (Silk \& Langer 2006). They are considered to be significant
contributors to various phenomena (Carr et al. 1984).

We have calculated the chemical enrichment evolution of the closed-box
system made only of VMOs. Marigo et al. (2003) give the evolutionary
properties of VMOs including lifetime, helium and metal yields, carbon and
helium cores at central C-ignition and mass loss for $120$--$1000M_{\odot}$.
Two different mass loss rates are considered by them:
the radiation-driven mass-loss model and the rotation-driven model.
The yields and mass loss rates of the two models are slightly different
from each other, and we use the rotation-driven model because it
produces a larger amount of  helium.
We find linear fits to these yields in the logarithmic scale
(Fig. 3) and show the resulting $\Delta$$Y$/$\Delta$$Z$
for each mass of VMO. The helium enrichment parameter of VMOs ranges
$63$ -- $6\times10^7$!

\begin{figure}
\begin{center}
\includegraphics[width=8cm]{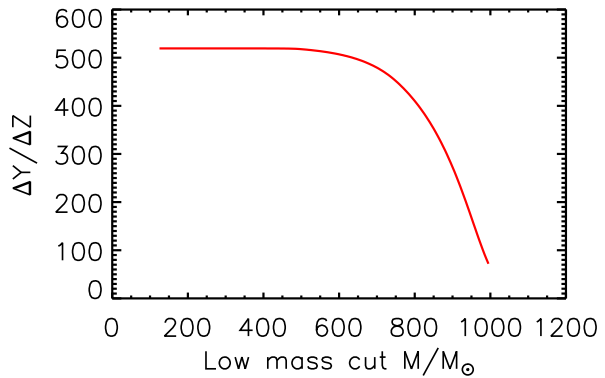}
\caption{The mass-cut dependence of $\Delta$$Y$/$\Delta$$Z$ of the
ejecta from VMOs. We constrain high mass cut of IMF as
1000 $M_{\odot}$ and IMF slope as canonical value, $x=1.35$. The system consisting of 120$\sim$1000$M_{\odot}$ VMOs
with Salpeter IMF slope results in $\Delta$Y/$\Delta$Z$\sim520$ and 995$\sim$1000$M_{\odot}$ system produces
$\Delta$Y/$\Delta$Z$\sim72$.}
\end{center}
\end{figure}

We have calculated $\Delta$$Y$/$\Delta$$Z$ for a population of VMOs
with various mass cuts and IMF slopes. The $\Delta$$Y$/$\Delta$$Z$ value after
the 1 Gyr evolution in a closed-box system varies with IMF slope and mass cut.
The values of $\Delta$$Y$/$\Delta$$Z$ as a function of lower mass cut are
shown in Fig 4. We fix the high mass cut as 1000$M_{\odot}$ assuming the
Salpeter IMF.
The resulting range of $\Delta$$Y$/$\Delta$$Z$ after 1Gyr of evolution
is two orders of magnitude higher than that of an ordinary system.
Our model matches the values of both $Y$ ($\sim 0.4$) and
$\Delta$$Y$/$\Delta$$Z$ ($\sim 70$) at age $\sim 1$~Gyr suggested for the
$bMS$ sub-population, as marked in Fig. 5.
Interestingly the solution suggests $M(VMO) \sim 995$--$1000M_{\odot}$;
but considering the uncertainty in the VMO yields, it may not be significant.

\begin{figure}
\begin{center}
\includegraphics[width=8cm]{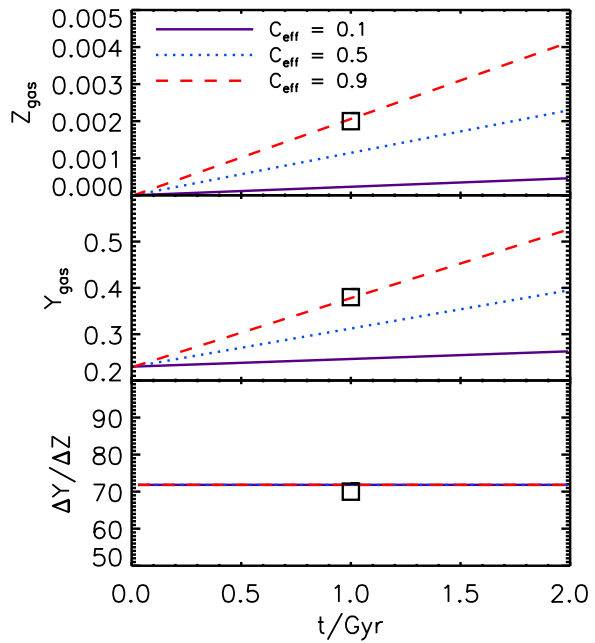}
\caption{
Chemical evolution in the gas of the VMO system 
for different values of star formatin efficiency. 
The values of helium and $\Delta$$Y$/$\Delta$$Z$ of the bMS population
of $\omega~Cen$ are marked ($squares$) in each panel.
\label{fig3}}
\end{center}
\end{figure}

\section{Discussion}

By means of a simple chemical evolution test, we conclude that ordinary
populations cannot produce $\Delta$$Y$/$\Delta$$Z$ much greater
than 4. None of the candidates succeeds in reproducing
the large helium enrichment claimed for the blue MS stars of $\omega$~Cen.
The reason for this is simple: the evolution of ordinary stars
result only in a modest value $\Delta$$Y$/$\Delta$$Z \sim 1$--5.
We confirm the result of Karakas et al. (2006) who rejected the AGB solution
on a similar ground.
The massive rotating star models of Maeder \& Meynet (2006) reach 
high values of $\Delta$$Y$/$\Delta$$Z$, but it is difficult
to maintain the high value longer than a few Myr in a $population$
with a realistic mass spread.

We present Pop III very massive objects (VMOs) as an alternative
solution. They produce both the helium abundance and the helium-to-metal
ratio searched for at the age $t \approx 1$ Gyr demanded 
by the observational constraint.
This scenario requires an overlap between the star formation episodes
of Pop~III and Pop II. Suppose that $\omega$~Cen formed in
the era of overlap of Pop~III and Pop II formation.
Earlier generations of Pop III stars form and their mass ejecta are
mixed in the short timescale of roughly $10^{7-8}$ yrs.
First generation of Pop~II stars start forming globally (this is not
part of $\omega$~Cen) and the processed gas gets recycled and mixed 
in the proto-system until the gas reaches $<$$Z$$> \sim 0.001$.
By now, the mixed $mean$ helium abundance is not any more extraordinary
but with {\em severe irregularities}.
The $rMS$ population of $\omega~Cen$ forms with the mean chemical composition.
A group of Pop III VMOs form in the tail of the Pop III episode nearby out
of a $pristine$ (irregularity) gas cloud.
Soon after this, that is, before their ejecta are mixed fully, the younger
stars of $\omega$~Cen ($bMS$) form under the heavy influence of this
helium-rich ejecta.
After the dynamical relaxation, the two chemically-distinct populations 
could be in one system. 
Since this is a geographic and chronological stochastic effect, it would not be
significant in the galactic scale.

The plausibility of our scenario strongly depends on the validity
of Marigo et al.'s yields. Other groups have computed the yields for
zero-metallicity VMOs (e.g. Bond et al. 1984; Ober et al. 1983; Klapp 1984),
and Marigo et al.'s yields are not in a smooth continuation with the yields 
of low-mass zero-metal stars given by Limongi \& Chieffi (2002).
The discontinuity may not be a big issue if the mass loss in VMOs may
happen in an extreme fasion (Smith 2006).
Considering the acute dependence of our conclusion on the reliability
of the chemical yields, it is urgent to perform detailed
independent stellar evolution modelling for VMOs.

The second question about this scenario is on the condition of VMO formation.
The Pop II star formation is allowed in the broad regions of
Pop~III objects (Tsujimoto et al. 1999; Susa \& Umemura 2006).
The $bMS$ population of $\omega$~Cen has $\sim 30$\% of the cluster
mass (Lee et al. 2005). Assuming $M_{tot} \sim 10^6~M_\odot$,
the helium-rich population mass would be $M_{bMS} \sim 3 \times 10^5~M_\odot$.
Assuming the primordial helium abundance is 23\%, 17\% 
of the $bMS$ population mass, $M_{He} \sim 5 \times 10^4~M_\odot$,
is the newly-generated helium mass.
To generate $5 \times 10^{4}M_{\odot}$ pure helium from 1000~$M_{\odot}$ VMOs,
we need at least $\sim 200$ VMOs simultaneously.
However, Abel, Bryan, \& Norman (2002) suggest that metal-free stars form
in isolation due to the immense radiation from the first-forming star.
Moreover, such a large number of Pop III stars forming close to each other,
if that is possible at all, would make the site extremely hostile for the
younger generation of $\omega$~Cen to form.
This poses serious challenges to our scenario.
By definition, this stochastic effect works only on small scales and not in
the galactic scale.
But the number of VMOs required would be smallest in the
deepest local gravitational potential, which means that this rare
event of achieving a high $\Delta$$Y$/$\Delta$$Z$, would prefer
larger gravitationally-bound objects, such as massive globular clusters
rather than small ones.

\section*{Acknowledgments}
We thank the anonymous referee for useful comments.
We are grateful to Ignacio Ferreras, Nobuo Arimoto, 
Volker Bromm, Greg Bryan, Brad Gibson,
Young-Wook Lee, Jason Tumlinson, and Suk-Jin Yoon for stimulating discussions.
This was supported by grant No. R01-2006-000-10716-0 from the Basic Research
Program of the Korea Science and Engineering Foundation and
by Yonsei University Research Fund of 2005 (S.K.Y.).


\end{document}